\documentclass{cmbe24} 
\usepackage{amsmath,amsfonts,amssymb}
\usepackage{graphicx,wrapfig}
\usepackage[hyphens]{xurl}
\usepackage{hyperref}

\usepackage{float,url,bm}

\newcommand{\bb}{\bm{b}}

\usepackage{amsthm}

\def\bx{\bm{x}}
\def\by{\bm{y}}

\def\bp{\bm{p}}

\newcommand{\bg}{\bm{g}}

\DeclareMathOperator*{\argmax}{arg\,max}
\DeclareMathOperator*{\argmin}{arg\,min}

\newcommand{\norm}[1]{\left\lVert#1\right\rVert}

\title{Data-driven imaging geometric recovery of ultrahigh resolution robotic micro-CT for in-vivo and other applications}

\author[1]{Mengzhou Li}
\author[2]{Guibin Zan}
\author[2]{Wenbin Yun}
\author[3]{Josef Uher}
\author[1]{John Wen}
\author[1]{Ge Wang}
\affil[1]{Biomedical Imaging Center, Rensselaer Polytechnic Institute, Troy, NY, USA \texttt{\{lim34,wenj,wangg6\}@rpi.edu}}
\affil[2]{Sigray Inc., Concord, CA, USA, \texttt{\{gbzan,wyun\}@sigray.com}}
\affil[3]{Radalytica a.s., U Pergamenky 1145/12, Prague, CZE, \texttt{josef.uher@radalytica.com}}

\summary{We introduce an ultrahigh-resolution (\(50\mu m\)) robotic micro-CT design for localized imaging of carotid plaques using robotic arms, cutting-edge detector, and machine learning technologies. To combat geometric error-induced artifacts in interior CT scans, we propose a data-driven geometry estimation method that maximizes the consistency between projection data and the reprojection counterparts of a reconstructed volume. Particularly, we use a normalized cross correlation metric to overcome the projection truncation effect. Our approach is validated on a robotic CT scan of a sacrificed mouse and a micro-CT phantom scan, both producing sharper images with finer details than that prior correction.

}

\keywords{Computed tomography, Robot arms, Imaging geometry 
estimation, Projection truncation}

\begin{document}

\section{Introduction}
Robot arm-based CT gains increasing traction in industrial applications and are now finding unique applications in biomedical imaging, especially for interior tomography. With the breakthrough in high-resolution (HR) photon-counting detectors (PCDs) and high-power micro-focus X-ray tubes, ultrahigh-resolution (UHR) interior CT becomes feasible for in vivo imaging with a most versatile scanner being developed using high-precision collaborative robots~\cite{li2020clinical}. Such an UHR micro-CT system would be particularly useful for various niche clinical applications, including plaque characterization, temporal bone imaging, lung nodule detection, bone modeling, and so on. However, as spatial resolution increases, the challenges of image artifacts become more pronounced, arising from patient movement, minor system misalignment, and robot coordination errors. 

Indeed, the drastically improved resolution requirement challenges the traditional assumptions about the smoothness of patient motion or geometric errors over projection views, which has underpinned most existing motion compensation methods~\cite{wicklein2013online,jang2019head}. In our prior work, we presented a data-driven method using locally linear embedding (LLE) principle to simultaneously correct all the geometric errors in an unified 9-degree-of-freedom framework~\cite{li2022motion2}. However, like the other existing methods, its performance is quite limited for interior CT scans which only collect bilaterally-truncated projections posing unique challenge for geometry estimation~\cite{sun2021motion}. Here we report our latest progress along this direction to upgrade our LLE-based geometry correction method using a normalized cross correlation (NCC) metric and enable region-of-interest (ROI) imaging of human carotid plaque at 50\(\mu m\) resolution with a robotic CT system.

\section{Methodology}
\subsection{UHR Robotic Interior CT Design for Carotid Plaque Imaging}
Heart attack and stroke have been leading causes of death worldwide, with disruption of vulnerable atherosclerotic plaques as a major etiology. Early detection and risk stratification of plaque vulnerability would be ideal for treatment planning and follow-up. Based on histology analysis, a key marker of high-risk plaques is a thinned fibrous cap (\(<65\mu m\)) with a lipid-rich necrotic core. However, a spatial resolution better than 65\(\mu m\) is required to measure the thickness objectively, which is beyond the capability of current clinical imaging methods, including CT, MRI, and ultrasound imaging (US). 

Here we propose a robotic arm-based UHR CT system for this purpose, integrating the latest breakthroughs in HR PCDs, ultrabright micro-focus X-ray tubes, modern robotics, and machine learning. Specifically, our system employs two collaborative robot arms, one equipped with a Sigray source (30\(\mu m\) focal spot, 150W power, 60-120kVp) and the other with an ADVACAM WidePIX1x5 PCD (55\(\mu m\)~detection resolution, \(70\times30 mm^2\) detection area), to optimize interior tomography of a targeted plaque. By always facing the source, along a trajectory, towards the ROI and adjusting the detector's position and orientation accordingly for a consistent magnification, we perform a patient-centered ROI scan to cut radiation dose and improve flux efficiency by minimizing the average source-ROI distance. With a \(2.67\times\) system magnification and \(2\times2\) pixel binning on the PCD, the calculated system resolution is 45\(\mu m\) over a maximum \(50mm\) field of view (via a proper detector offset during the scan). When mimicking the human neck with a \(13cm\) water slab, we expect to receive over 4,000 photons per \(110\mu m\) pixel under 0.1 second exposure and \(342.4mm\) source-to-detector distance at the maximum tube power, allowing 360 quality projections within 36 seconds for image reconstruction. 

As a potential extension, we could integrate a US probe on one robotic arm to automate a US scan and reconstruct a US volume from the 2D B-scan series. This volumetric US imager would also provide blood flow information. The X-ray and US hybrid imaging capability will significantly enhance the screening/diagnostic imaging performance.

\subsection{LLE-based Geometry Estimation for Interior Scans}
\textbf{Tomographic Consistency} Following the work~\cite{li2022motion2}, the data-driven geometry estimation based on tomographic consistency is formulated as an optimization problem:
\vspace{-2truemm}
\begin{equation}
    \vspace{-2truemm}
    \argmin_{\bg,\bx} \norm{\bb - A(\bg)\bx}_L, \label{eq:TomoConsistency}
\end{equation}
where \(A(\cdot)\) denotes the system matrix as a function of the geometry parameters \(\bg=[\bg_1,\cdots,\bg_i,\cdots,\allowbreak \bg_{N_v}]^T\), \(N_v\) is the total number of views, \(\bx\) and \(\bb\) are the volumetric reconstruction and projection data respectively, and the mismatch is measured by the metric \(L\). Usually, the \(L_2\)-norm is used. We alternately update the motion estimation \(\bg\) and image reconstruction \(\bx\) to solve the problem.
As the motion parameters for different views are independent, we estimate the parameters view-by-view. One sub-problem for a typical view is expressed as
\vspace{-2truemm}
\begin{equation}
  \bg_i^\prime = \argmin_{\bg_i} \norm{\bb_i - A_i(\bg_i)\bx^\prime}_L,\label{eq:MotionEst}
  \vspace{-2truemm}
\end{equation}
where \(\bx^\prime\) is the latest reconstruction, and \(\bg_i^\prime\) is the parameter vector to be estimated for view \(i\).

\textbf{LLE-based Optimization} Different from gradient descend optimization to solve Eq.~\ref{eq:MotionEst}, LLE densely samples a pre-defined parametric space and directly infers the solution from its nearest sampled points according to embedding theory~\cite{li2022motion}. The principle is that given a dense enough sampling grid, the true motion parameter \(\bg_i^*\) should be sufficiently close to its \(K\)-nearest neighbors in the grid such that both the low-dimensional motion parameter \(\bg_i^*\) and corresponding high dimensional projection \(\bb_i\) can be linearly embedded by their neighbors, and both embeddings share the same weights, i.e.,
\vspace{-3truemm}
\begin{equation}
  \bg_i^* = \sum_{k=1}^{K} w_{k}\bg_i^{(k)}, \quad \bb_i = \sum_{k=1}^{K} w_{k} \bp_i^{(k)} \quad \text{s.t.} \quad \sum_{k=1}^K w_{k} = 1, \label{eq:LLE2}
  \vspace{-2truemm}
\end{equation}
where \(\bg_i^{(k)}\) is one of the \(K\) nearest samples for \(\bg_i^*\) and \(\bp_i^{(k)} = A_i(\bg_i^{(k)})\bx^\prime\) is the corresponding reprojection. Thus, we may update the estimation \(\bg_{i}^\prime\) for \(\bg_i^*\) by solving the embedding weights \(\tilde{\bm{w}}\) on the \(K\) neighbors \(\{(\tilde{\bg}_i^{(k)}, \tilde{\bp}_{i}^{(k)}) | k = 1,\cdots,K\}\):
\vspace{-3truemm}
        \begin{equation}
          \bg_{i}^\prime = \sum_{k=1}^K{\tilde{w}_k\tilde{\bg}_{i}^{(k)}}, \quad \tilde{\bm{w}} = \argmin_{\bm{w}} \norm{\bb_i - \sum_{k=1}^K{w_k \tilde{\bp}_i^{(k)}}}_L \, \text{s.t.} \, \sum_{k=1}^K{w_k} = 1.  \label{eq:embeding}
          \vspace{-3truemm}
        \end{equation}

\textbf{{Normalized Cross Correlation-based Embedding}}
Despite the success of the L2 metric for regular CT scans~\cite{li2022motion,sun2016iterative}, it is less effective in truncated CT scans. To address the issue, we propose a normalized cross correlation-based embedding method. With a little abuse of notations without losing clarity, let us suppose that we have \(K\) neighbors \(\{\by_{j}|j = 1,\cdots, K\}\) to embed the ground truth projection \(\by_{*}\) with adjustable weighting coefficients \(\bm{w}=[w_1,\cdots,w_j,\cdots,w_K]^T\) to maximize the normalized cross correlation between the embedding \(\tilde{\by}\) and the ground truth \(\by_*\). Mathematically, we reformulate Eq.~\ref{eq:embeding} as the following optimization problem:
\vspace{-3truemm}
\begin{align}
             & \argmax_{\bm{w}} \quad \frac{(\tilde{\by}^T - \tilde{\mu})(\by_{*} - \mu_{*})}{\tilde{s}s_{*}} \quad\label{eq:NCC}               
  s.t. \quad \tilde{\by} = [\by_1,\cdots,\by_K]\bm{w}, \;\sum_{j=1}^{K}{w_j} = 1,                                                               \\
             & \tilde{\mu} = \sum_{i=1}^{N}{\tilde{\by}_{[i]}/N}, \tilde{s} = \sqrt{\sum_{i=1}^{N}{(\tilde{\by}_{[i]} - \tilde{\mu})^2/(N-1)}},
\end{align}
where \(\mu_*, s_*\) and \(\tilde{\mu},\tilde{s}\) are the mean and the standard deviation values of the ground truth \(\by_*\) and its embedded representation \(\tilde{\by}\), and \(\by_{[i]}\) indexes elements of \(\by\), and \(N\) is the total number of the elements. Due to limited space, we present the solution for the above optimization problem without intermediate derivation as follows:
\vspace{-3truemm}
\begin{equation}
  \bm{w} = \frac{[c_1/s_1,\cdots,c_K/s_K]^T}{\sum_{j=1}^{K}{c_j/s_j}}.
  \vspace{-3truemm}
\end{equation}

\section{Results and Discussions}
\begin{figure}
  \centering
  \includegraphics[width=0.95\linewidth]{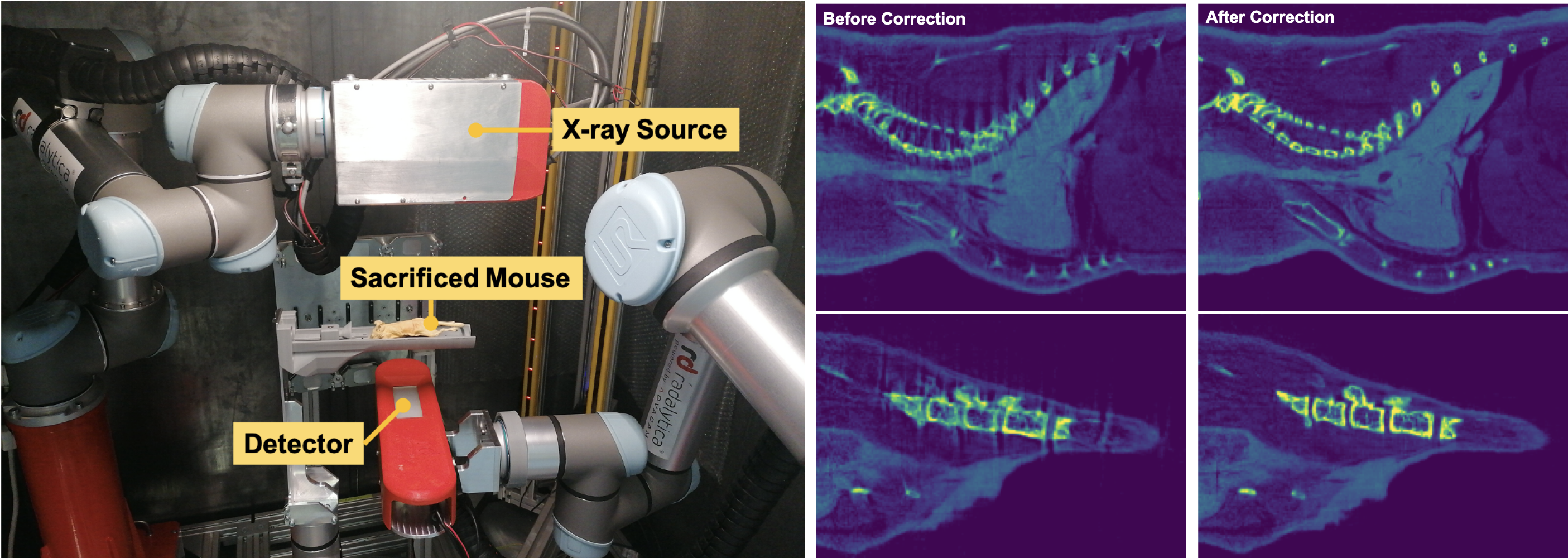}
  \vspace{-3truemm}
  \caption{Illustration of an exemplary robotic CT system scanning an euthanized mouse (left) and the zoomed ROI in a sagittal slice before and after our data-driven geometrical error correction (right).}\label{fig:Robots}
  \vspace{-5truemm}
\end{figure}

\begin{figure}
  \centering
  \includegraphics[width=0.95\linewidth]{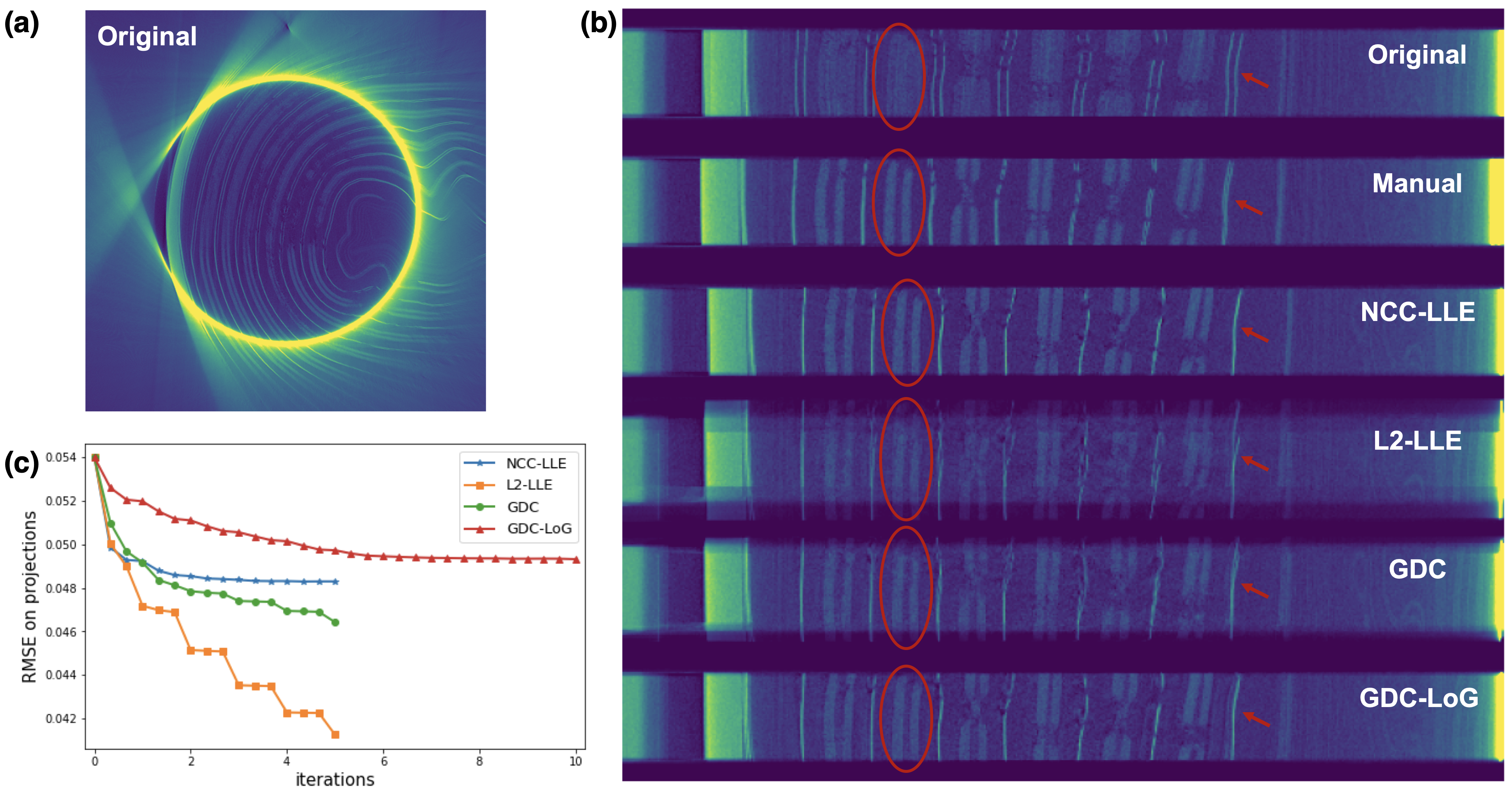}
  \vspace{-5truemm}
  \caption{Pantom study with and without geometric correction. (a) The field of view reconstruction of the battery phantom before correction; (b) zoom-in sagittal images of the region of interest before and after correction using the the manual alignment (Manual), NCC-LLE (ours), L2-LLE, GDC, and GDC-LoG methods respectively. The circles and arrows respectively indicate the positive electrode and negative current collector; and (c) root mean squared error (RMSE) curves between reprojections and real projections in correction iterations.}\label{fig:BatterySag}
  \vspace{-5truemm}
\end{figure}

We validated our method on (1) a robotic CT scan of an euthanized mouse with unknown misalignment and coordination errors and (2) a micro-CT scan of a lithium-ion battery from AirPods with unknown detector drifting. As shown in Fig.~\ref{fig:Robots}, sharper images with clearer anatomical details were obtained after our geometric calibration, showing the effectiveness of our method. In the battery phantom case, the regular structure can help us identify image artifacts and judge the motion compensation quality. As shown in Fig.~\ref{fig:BatterySag}, we compared our method with our previous method L2-LLE~\cite{li2022motion2}, the gradient consistency method (GDC)~\cite{sun2016iterative}, and the latest GDC-LoG method developed for dental CT scans with data truncations~\cite{sun2021motion}.
Figure~\ref{fig:BatterySag}(a) shows the direct reconstruction result without correction. The bright yellow ring represents strong truncation artifacts caused by the lateral projection truncation as the detector can only cover a portion of the battery. Serious double-edge artifacts are observed because of the detector drifting. 
Figure~\ref{fig:BatterySag}(b) shows the sagittal view of the region of interest (within the yellow ring). The circle and arrow indicate the positive electrode (thick structures) and negative current collector (a thin line) respectively. Clearly, our method gives the best contrast and the least residual double-edge artifacts. The residual double-edge artifacts are visible for manual alignment though improved relative to the original before correction.The L2-LLE and GDC methods present ghost images at the top and bottom while the other methods do not. Since the original reconstruction does not present similar shadows, this suggests that the ghosts may be produced by L2-LLE and GDC due to the insufficiency of the L2 metric for truncated data. The GDC-LoG method achieves the second best image quality despite some residual artifacts near the arrow in the sagittal view, suggesting the effectiveness of the LoG operation in improving the L2 metric for tomographic consistency.
Figure~\ref{fig:BatterySag}(c) presents the root mean squared error measured in the projection domain through iterations. Despite the bad image quality of GDC and L2-LLE methods, they reach significantly smaller RMSE loss than the L2-LLE and GDC-LoG methods. This suggests that the L2 metric may not be the most effective to measure the consistency for interior tomography due to the associated truncation artifacts. Furthermore, our method demonstrates faster convergence (1.25 hours for 2 iterations) than the GDC-LoG method (2.72 hours for 6 iterations). 

In conclusion, we have presented an UHR robotic CT design for ROI imaging of carotid plaques and a corresponding NCC-based embedding optimization method for geometrical calibration of interior scanning geometry. Our LLE framework does not rely on any trajectory smoothness assumption and is particularly useful for high-resolution CT imaging. An analytic solution to the NCC optimization problem has been derived to boost the computational efficiency. The experimental results have demonstrated great advantages of our method over the competing methods in terms of both accuracy and efficiency, and a great potential of our approach for UHR robotic interior tomography.

\textbf{Acknowledgment} This work is supported in part by the NIH grant R01CA237267.

{\small
\bibliographystyle{IEEEtran}

}

\end{document}